\documentclass[prb,twocolumn,a4paper,showpacs,citeautoscript,floatfix]{revtex4}

\usepackage{amsfonts,amsmath,amssymb}
\usepackage{graphicx}
\usepackage{bm}

\newif\iffullbib
\fullbibtrue

\begin{document}

\title{Preeminent role of the Van Hove singularity in the strong-coupling
analysis of scanning tunneling spectroscopy for two-dimensional cuprates}

\author{Giorgio Levy de Castro}
\email{giorgio.levy@physics.unige.ch}
\author{Christophe Berthod}
\author{Alexandre Piriou}
\author{Enrico Giannini}
\author{{\O}ystein Fischer}
\affiliation{DPMC-MaNEP, Universit\'e de Gen\`eve, 24 quai
Ernest-Ansermet, 1211 Gen\`eve 4, Switzerland}

\date{\today}

\begin{abstract}

In two dimensions the non-interacting density of states displays a
Van Hove singularity (VHS) which introduces an intrinsic
electron-hole asymmetry, absent in three dimensions. We show that
due to this VHS the strong-coupling analysis of tunneling spectra
in high-$T_c$ superconductors must be reconsidered. Based on a
microscopic model which reproduces the experimental data with
great accuracy, we elucidate the peculiar role played by the VHS
in shaping the tunneling spectra, and show that more conventional
analyses of strong-coupling effects can lead to severe errors.

\end{abstract}

\maketitle

Scanning tunneling spectroscopy of Bi-based cuprate high-$T_c$ superconductors
(HTS) shows a $d$-wave gap and a strong dip-hump feature which is nearly always
stronger for occupied than for empty states \cite{Renner-1995}. It has been
proposed that the dip-hump structure results from the interaction of electrons
with a collective mode \cite{Campuzano-1999}, but the dip asymmetry has not
received an explanation so far. Indeed such a coupling leads to electron-hole
symmetric spectra in classical superconductors \cite{Eliashberg-1960,
Giaever-1962, McMillan-1965}. The dip-hump was also observed by photoemission,
but in those experiments it is not possible to probe the electron-hole
asymmetry. Sometimes photoemission spectra are even symmetrized
\cite{Campuzano-2004}, thus ignoring the relevance of the asymmetry seen in
tunneling. The fact that in two dimensions the density of states (DOS) has a
prominent van-Hove singularity (VHS), unlike in 3D, introduces naturally an
asymmetry and thereby modifies the strong-coupling analysis and the
corresponding determination of the collective mode frequency in an essential
way.

Photoemission experiments have provided a detailed account of the band structure
in cuprates \cite{Damascelli-2003, Feng-2002, Sato-2003}. In agreement with
early calculations \cite{Pickett-1989}, the band crossing the Fermi level
presents a saddle point leading to a logarithmic VHS in the DOS. The scanning
tunneling microscope (STM) is the ideal tool to look for such singularities,
since under suitable conditions it probes directly the DOS with meV resolution
\cite{Tersoff-1983, Chen-1990a, Fischer-2006}. Up to now, however, there has
been no report of a direct STM observation of the VHS in HTS materials, neither
in the normal nor in the superconducting state. Previous interpretations of the
missing VHS invoked the tunneling matrix element: \cite{Kouznetsov-1996,
Yusof-1998} in planar junctions, it is indeed believed that the DOS features in
the direction normal to the junction are hidden due to a cancellation with the
electron velocity, and that the DOS in the plane of the junction does not show
up due to focalization effects \cite{Harrison-1961}. These two mechanisms cannot
explain the absence of VHS in $c$-axis STM/HTS tunnel junctions: the HTS
materials being quasi two-dimensional have virtually no dispersion in the
tunneling direction, and the STM junction, owing to its microscopic size, is
qualitatively different from a planar junction and is characterized by a
specific matrix element which may not lead to focalization effects
\cite{Tersoff-1983, Chen-1990a}. Besides, the ability of the STM to probe DOS
singularities was recently demonstrated in carbon nanotubes \cite{Ouyang-2001}.

The solution to this puzzle lies in the coupling to collective modes. Apart from
inducing the dip feature \cite{Eschrig-2000, Zasadzinski-2001}, this coupling
was also shown to effectively suppress the VHS peak in STM spectra
\cite{Hoogenboom-2003b}. Here we demonstrate that the interplay of the VHS and
the collective mode concludes the picture, providing a complete explanation of
both the missing VHS and the pronounced electron-hole asymmetries. Our
conclusions are based on a brute force fit of the STM spectra and a careful
analysis of the model. The advantage of this method is to allow for unambiguous
determinations of relevant physical parameters from the raw tunneling data, and
to provide intuition about the relationship between trends in the spectral
features and parameter variations.

\begin{figure}[t]
\includegraphics[width=6.5cm]{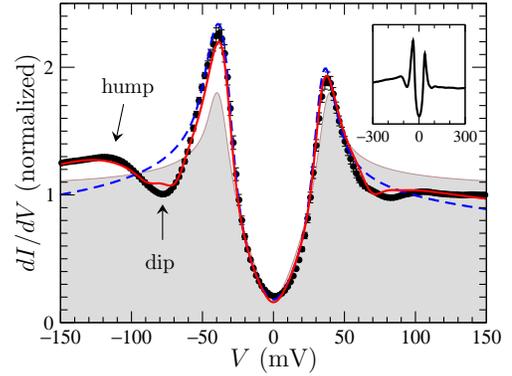}
\caption{\label{fig1} Typical STM conductance of Bi2223
($T_c=111$~K) at $T=2$~K (dots). The data is an average of several
spectra taken at different positions on the same sample, and
having the same peak-to-peak gap $\Delta_p=38$~meV. Error bars
give the standard deviation of this average. The inset shows the
spectrum on a larger energy scale. Also shown are three model
predictions (see text): free-electrons $d$-wave BCS (shaded area),
$d$-wave BCS with realistic dispersion including VHS (dashed
line), and $d$-wave BCS including VHS and coupling to the
collective mode (full line). The total spectral weight in the
energy range of the figure is the same for all curves. }
\end{figure}

As a template material we studied the three-layer compound
Bi$_2$Sr$_2$Ca$_2$Cu$_3$O$_{8+\delta}$ (Bi2223), which has the highest $T_c$ in
the Bi-based family. Reproducible spectroscopy was recently reported in single
crystals of Bi2223 cleaved \textit{in situ} at room temperature
\cite{Kugler-2006}. The tunneling conductance measured in UHV on an
optimally-doped sample is shown in Fig.~\ref{fig1}. The spectrum presents the
characteristic V shape of $d$-wave superconductors, strong and asymmetric
coherence peaks, and an asymmetric dip-hump structure. We have deliberately
selected an optimally-doped sample displaying a flat background conductance up
to high energies (inset in Fig.~\ref{fig1}). For underdoped samples the spectra
acquire an asymmetric background, attributed to strong correlation effects
\cite{Rantner-2000, Randeria-2005, Anderson-2006}, which can conceal other
intrinsic asymmetries. By moving to optimal doping where this background is
absent, we can thus exclude that the observed asymmetries result from this type
of correlations.

For illustrative purposes, we plot in Fig.~\ref{fig1} the prediction of a
conventional free-electrons BCS $d$-wave model (curve defined by the shaded
area) \cite{Won-1994, Smearing}. This model fits the experimental data well at
low energy ($\lesssim\Delta_p/2$), but fails to account for the various features
present at higher energy, in particular the asymmetry of the coherence peak
height. A much better description of the coherence peak height, width, and
asymmetry can be achieved by taking into account the actual band structure. At
energies below $\sim200$~meV a one-band model turns out to be sufficient. We
consider the two-dimensional lattice model $\xi_{\bm{k}}=2t_1(\cos k_x+\cos
k_y)+4t_2\cos k_x\cos k_y+2t_3(\cos2k_x+\cos2k_y)-\mu$ where $t_i$ is the $i$-th
neighbor hopping energy. For this dispersion the VHS lies at energy
$\xi_{\text{M}}=-4(t_2-t_3)-\mu$ corresponding to the saddle point
$\bm{k}=(\pi,0)\equiv\text{M}$. We determined the parameters of the band through
an unconstrained least-squares fit of the whole spectrum in the inset of
Fig.~\ref{fig1}, leading to $t_1=-882$, $t_2=239$, $t_3=-14$, and
$\xi_{\text{M}}=-26$~meV, as well as a $d$-wave gap $\Delta_0=34.1$~meV. It is
very encouraging that these numbers determined from nothing else than the STM
tunneling conductance lead to a Fermi surface in semi-quantitative agreement
with the one measured by photoemission \cite{Feng-2002}. In particular, the VHS
lies below the Fermi energy as expected for a hole-doped material. The resulting
theoretical curve (Fig.~\ref{fig1}, dashed line) is very similar to the
free-electron model at subgap energies, but performs much better up to an energy
slightly above the coherence peaks. The main effect of the VHS is to provide
additional spectral weight below the Fermi level and thus increase the height of
the coherence peaks at negative bias. Note that in the absence of broadening
factors \cite{Smearing} the VHS would be visible as a secondary peak flanking
the main coherence peak (see Fig.~\ref{fig2} below).

The ``BCS plus VHS'' model is nevertheless not satisfactory above
$eV\sim2\Delta_p$, where it fails to reproduce the significant transfer of
spectral weight from the dip to the hump, which is strongest at negative bias in
the experimental spectrum. Generically, such transfers signal a strong coupling
of the quasiparticles with a collective excitation, which leads to enhanced
damping of the former in a limited energy range, and to a simultaneous
renormalization of the dispersion. In conventional superconductors, the
electron-phonon coupling is known to induce similar features, albeit much less
pronounced, at biases related to the phonon frequencies \cite{Eliashberg-1960,
Giaever-1962, McMillan-1965}. A phonon-based interpretation of the dip-hump in
HTS has been revivified recently \cite{Sandvik-2004, Devereaux-2004, Citro-2006,
Lee-2006a, Pilgram-2006, Zhao-2006}. Another candidate is the famous $(\pi,\pi)$
magnetic excitation \cite{Campuzano-1999, Eschrig-2000, Zasadzinski-2001}
commonly known as the ``41~meV resonance''---although its energy changes from
one material to the other and also with varying doping \cite{Sidis-2004}.
Coupling the quasiparticles to this collective mode yields a change of the
electron self-energy which can be expressed to leading order in terms of the
spin susceptibility $\chi_s(\bm{q},\omega)$ \cite{Eschrig-2000}. Using a
parametrization of $\chi_s$ as measured by inelastic neutron scattering,
Hoogenboom \textit{et al.} showed that this model provides a very good
description of the STM spectra of Bi2212 at several dopings
\cite{Hoogenboom-2003b}. Apart from the band-structure parameters $t_i$, $\mu$
and the $d$-wave gap $\Delta_0$, this model has 3 more parameters, namely the
resonance energy $\Omega_s$, a characteristic length $\xi_s\sim2a$ which
describes the spread of the collective mode around $\bm{q}=(\pi,\pi)$
\cite{Eschrig-2000}, and a coupling constant $g$.

In order to estimate these parameters we again performed a least-squares fit of
the whole spectrum in the inset of Fig.~\ref{fig1}, however keeping the $t_i$'s
fixed to their values determined previously. This procedure yields
$\Delta_0=33.9$~meV and $\xi_{\text{M}}=-42.4$~meV, as well as
$\Omega_s=34.4$~meV, in reasonable agreement with the properties of the magnetic
resonance measured in Bi2223 \cite{Pailhes}. The resulting theoretical spectrum
matches our experimental data with amazing accuracy (Fig.~\ref{fig1}, full
line). In particular, the model reproduces all of the asymmetries found
experimentally between positive and negative biases. We would like to stress
that these asymmetries cannot be understood in models which neglect the band
structure, as, e.g., in Refs~~\onlinecite{Pilgram-2006} and~
\onlinecite{Zasadzinski-2006}. The shape of the dip minimum in the theory
differs somewhat from experiment: we shall come back to this below. Fits of
similar quality have been obtained for many different spectra with gaps varying
from $\Delta_p=36$ to 54~meV.

The precise interpretation of the theoretical curve in Fig.~\ref{fig1} seems
complicated due to the interplay of three similar energy scales: the $d$-wave
gap $\Delta_0$, the VHS energy $\xi_{\text{M}}$, and the collective mode energy
$\Omega_s$, all in the 30--40~meV range. Still, based on a careful study of the
model we can identify the origin of each structure in the spectrum, as
illustrated in Fig.~\ref{fig2}. The bare BCS DOS $N_0(\omega)$ exhibits 5
singularities, namely ($a$) the V at zero energy resulting from the $d$-wave
gap; ($b$) and ($b'$) the coherence peaks at negative and positive energies
($-\omega_b$ and $\omega_{b'}$ respectively); ($c$) the VHS at energy
$-\omega_c$ below the coherence peak, and ($c'$) the weak echo of the VHS at
energy $\omega_{c'}$ due to the BCS electron-hole mixing. The interaction with a
collective mode leads to inelastic processes in which a quasiparticle of
momentum $\bm{k}$ and energy $\omega$ is scattered to a state with momentum
$\bm{k}-\bm{q}$ and energy $\omega-\Omega$ through emission of a collective
excitation with quantum numbers $(\bm{q},\Omega)$. The corresponding self-energy
diagram is sketched in Fig.~\ref{fig2}. If the only excitation available is a
sharp-in-energy mode, all singularities of $N_0(\omega)$ are mirrored in the
self-energy, and exactly shifted by the mode energy $\Omega_s$. Hence the DOS
$N(\omega)$ including the interaction with the mode displays 3 pairs of
singularities indicated by arrows in Fig.~\ref{fig2}: the onsets at
$\omega=\pm\Omega_s$, below which the quasiparticles do not have enough energy
to excite a collective mode, a first minimum in the dip at $-\omega_b-\Omega_s$
(resp. $\omega_{b'}+\Omega_s$) corresponding to the negative-energy
(positive-energy) coherence peak, and a second minimum in the dip---echoing the
VHS peak in $N_0(\omega)$---which is more pronounced for occupied states at
$-\omega_c-\Omega_s$, but also visible at $\omega_{c'}+\Omega_s$. Therefore the
asymmetry of the dip structure between positive and negative biases receives a
natural explanation in terms of the asymmetry of the underlying BCS DOS, which
in turn is due to the VHS. The appearance of a double minimum in the dip is a
direct consequence of the BCS DOS having both a coherence peak at $-\omega_b$
and a VHS peak at $-\omega_c$. Such a double minimum is not observed in the
experimental spectrum of Fig.~\ref{fig1}. At positive bias, the various
broadening effects \cite{Smearing} are sufficient to smear out the two minima
into one. On the other hand, we have found that if the collective mode has a
finite inverse lifetime of only $\sim6$~meV \cite{Pailhes}, then the two minima
in the dip fade away resulting in a smooth dip also at negative bias as observed
experimentally.

\begin{figure}[tb]
\includegraphics[width=8cm]{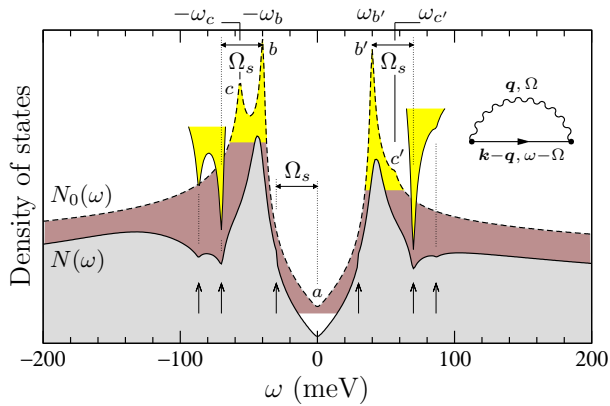}
\caption{\label{fig2}
High-resolution calculation of the DOS in the absence [$N_0(\omega)$, dashed
line, shifted vertically] and in the presence [$N(\omega)$, full line] of the
coupling to the $(\pi,\pi)$ resonance. The energy and shape of the structures
induced by this coupling (arrows) is an ``inverted image'' of the singularities
present in the BCS DOS ($a$, $b$, $b'$, $c$, $c'$), shifted by the mode energy
$\Omega_s$. The parameters are as in Fig.~\ref{fig1}, except for $\Delta_0=40$,
$\xi_{\text{M}}=-40$, and $\Omega_s=30$~meV. The inset shows the self-energy
diagram with the full line representing the BCS Green's function and the wavy
line the spin susceptibility.
}
\end{figure}

\begin{figure}[tb]
\includegraphics[width=8.5cm]{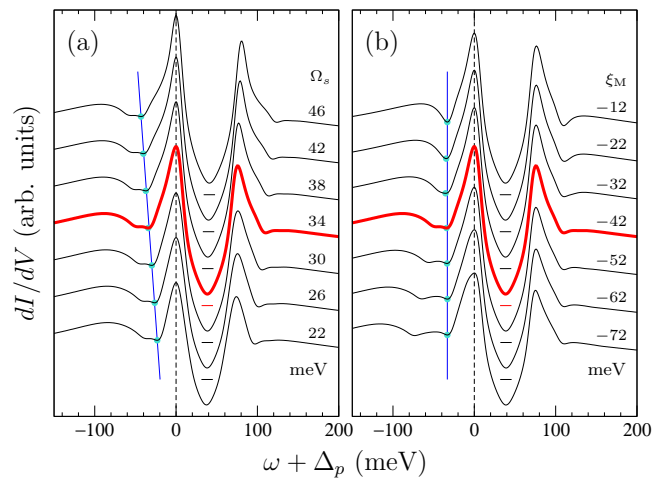}
\caption{\label{fig3}
Evolution of the theoretical tunneling conductance upon varying $\Omega_s$ (a)
or $\xi_{\text{M}}$ (b). The other model parameters are as in Fig.~\ref{fig1}.
Energies are measured relative to the peak maximum at $-\Delta_p$. The values of
$\Omega_s$ and $\xi_{\text{M}}$ span a larger range than their possible
variation in the experiment (Fig.~\ref{fig4}), but were chosen so with the aim
of amplifying the trends.
}
\end{figure}

The exact relationship between the position of the various structures in
$N(\omega)$ and the parameters of the model is not straightforward. The peak
maximum, which we have denoted by $\Delta_p$, is not related in any simple way
to the structures in $N_0(\omega)$, although it is numerically close to
$\omega_b$. Furthermore $\omega_b$ is not given by $\Delta_0$, but by the value
of the gap function $\Delta(\bm{k})=\frac{\Delta_0}{2}(\cos k_x-\cos k_y)$ at
the Fermi crossing near the M point. Hence $\omega_b$ is slightly smaller than
$\Delta_0$. The case of $\omega_c$ is simpler, and it can be shown that
$\omega_c=\sqrt{\xi_{\text{M}}^2+\Delta_0^2}$. It follows that the first minimum
in the dip at negative energy lies to a good approximation at
$-\Delta_p-\Omega_s$, and the second at
$-\sqrt{\xi_{\text{M}}^2+\Delta_p^2}-\Omega_s$. In Fig.~\ref{fig3} we illustrate
these two dependencies by varying $\Omega_s$ and $\xi_{\text{M}}$ independently
in the model. Our starting point is the spectrum of Fig.~\ref{fig1} reproduced
in bold in Fig.~\ref{fig3}. In order to facilitate the comparison we have
positioned the spectra relative to the negative-energy coherence peak. Varying
$\Omega_s$ while keeping $\xi_{\text{M}}$ fixed we clearly see that the main
change in the spectrum is a displacement of the dip and hump relative to the
peak, consistently with the interpretation given in Fig.~\ref{fig2}. In
particular, the width of the dip at negative bias does not depend on $\Omega_s$.
As $\Omega_s$ increases, we also observe that the coherence peaks become taller
and thinner, while developing a shoulder. This shoulder carries part of the
spectral weight expelled from the dip, and progressively exits the coherence
peak as the difference in the energy scales $\Omega_s$ and $\Delta_0$ increases.
Fig.~\ref{fig3}(a) further shows that $\Omega_s$ has not much influence on the
electron-hole asymmetry of the spectra. In contrast, changing the position of
the VHS by varying $\xi_{\text{M}}$ dramatically affects this asymmetry. At the
lowest $\xi_{\text{M}}$ considered the spectrum is almost symmetric. As the VHS
moves toward negative energy, the dip at $\omega<0$ gets wider (the first
minimum in the dip does not move, as expected) and the dip at $\omega>0$ dies
out. Inspection of Fig.~\ref{fig3} also shows that the maximum of the hump
feature tracks the second minimum in the dip, and thus depends on both
$\Omega_s$ and $\xi_{\text{M}}$. Furthermore the hump gets flattened as
$|\xi_{\text{M}}|$ increases.

\begin{figure}[tb]
\includegraphics[width=5.5cm]{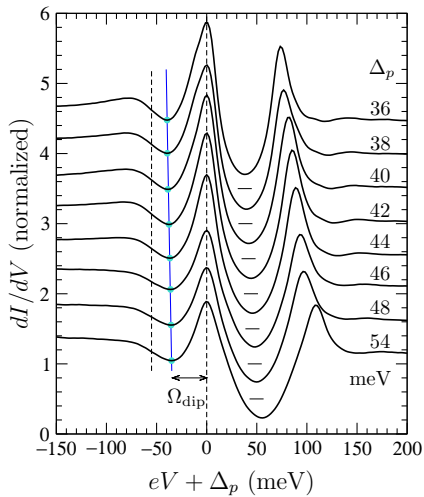}
\caption{\label{fig4}
STM conductance spectra of Bi2223 ($T_c=111$~K) at $T=2$~K. Each curve is an
average of several spectra taken at different locations on the same sample, all
having the indicated peak-to-peak gap $\Delta_p$. $\Omega_{\text{dip}}$ is the
energy difference between the dip minimum (dot) and the peak maximum at negative
bias, relative to which voltages are measured.
}
\end{figure}

Having clarified the different roles of $\Omega_s$ and $\xi_{\text{M}}$ in the
shape of the model DOS, we now come back to experiment. In Fig.~\ref{fig4} we
plot a series of tunneling conductance spectra with peak-to-peak gaps ranging
from $\Delta_p=36$ to 54~meV. As a function of increasing $\Delta_p$, we first
observe that (i) the dip for occupied states gets wider, (ii) the dip for empty
states gets weaker, and (iii) the hump at negative energy flattens out. These
three trends are all observed in Fig.~\ref{fig3}(b), and therefore strongly
suggest that the increase of $\Delta_p$ goes together with a displacement of the
VHS towards negative energies. This is also fully consistent with the idea that
local increases of $\Delta_p$ in inhomogeneous samples reflect local decreases
in the hole concentration. Another obvious trend of the data in Fig.~\ref{fig4}
is that the coherence peaks are reduced with increasing gap. As seen in
Fig.~\ref{fig3}(b), this is also consistent with a shift of the VHS to lower
energy. However, a look at Fig.~\ref{fig3}(a) shows that this trend can also be
ascribed to a decrease in the value of $\Omega_s$. Our calculations indeed
confirm that $\Omega_s$, as determined by fits to the spectra in
Fig.~\ref{fig4}, decreases from 34 to 24~meV with increasing $\Delta_p$. The
energy difference between the coherence peak and the dip minimum in the
experimental spectra, $\Omega_{\text{dip}}$, also decreases with increasing
$\Delta_p$, but less than $\Omega_s$ (from 39 to 35~meV), as can be seen in
Fig.~\ref{fig4}. From these numbers it appears clearly that
$\Omega_{\text{dip}}$ overestimates $\Omega_s$ by 5 to 10~meV. Recently the
energy difference between $\Delta_p$ and the inflection point between the dip
and the hump (extremum in the $d^2I/dV^2$ spectrum) was used as an estimate of
$\Omega_s$ in Bi2212 \cite{Lee-2006a}, resulting in an average value of 52~meV.
This same estimate would give a $\Delta_p$-independent result of $\sim57$~meV
for the data in Fig.~\ref{fig4}, almost a factor of two larger that $\Omega_s$.

In summary, we have shown that the van Hove singularity plays a crucial role in
shaping the spectral features induced in the STM spectra by the interaction of
quasiparticles with a collective mode. As a result, determining the frequency of
the mode from STM data is complicated, and cannot be done directly from simple
structures in the $dI/dV$ or $d^2I/dV^2$ spectra. This conclusion, obtained for
a coupling to the $(\pi,\pi)$ spin resonance, also applies to phonon models.

We thank I. Maggio-Aprile and Y. Fasano for useful discussions. This work was
supported by the Swiss Natio\-nal Science Foundation through Division II and
MaNEP.

\iffullbib

\else

\fi


\begin{thebibliography}{35}
\bibitem[{Sme()}]{Smearing}
\bibinfo{note}{Our theoretical curves in Figs~\ref{fig1} and \ref{fig3} show
  the DOS corrected by three broadening effects: a constant scattering rate
  $\Gamma=2$~meV, a $2$~K thermal smearing, and a gaussian broadening of
  $4$~meV. The latter accounts for the combined experimental uncertainty due to
  the lock-in ac modulation at $V_{\text{ac}}=2$~mV, and the averaging of
  several slightly different spectra.}
\end{thebibliography}
\end{document}